%% file: main.tex
\documentclass[conference]{IEEEtran}
\usepackage{glossaries}
\usepackage{color}
\usepackage{cite}
\usepackage{booktabs}
\usepackage{hyperref}
\usepackage{gensymb}
\usepackage{url}
\usepackage{enumitem}
\ifCLASSINFOpdf
   \usepackage[pdftex]{graphicx}

\else

\fi
\usepackage{amsmath}
\DeclareMathOperator{\atantwo}{atan2}

\usepackage{algorithmic}
\usepackage{algorithm}
\usepackage[printonlyused]{acronym} 
\acrodef{AoA}{angle-of-arrival}
\usepackage[caption=false,font=footnotesize]{subfig}
\usepackage{url}
\usepackage{amssymb}
\usepackage{bm}
\input{acronyms.tex} 
\usepackage{caption}
\begin{document}
\title{Cooperative Localization with Angular Measurements and Posterior Linearization}

\author{Yibo~Wu\IEEEauthorrefmark{1},
        Bile~Peng\IEEEauthorrefmark{1},
        Henk~Wymeersch\IEEEauthorrefmark{1},
        Gonzalo Seco-Granados\IEEEauthorrefmark{3},\\
        Anastasios~Kakkavas\IEEEauthorrefmark{2}\IEEEauthorrefmark{4},
        Mario H. Casta\~{n}eda Garcia\IEEEauthorrefmark{2},
        and Richard A. Stirling-Gallacher\IEEEauthorrefmark{2}\\
        \IEEEauthorrefmark{1}Department of Electrical Engineering, Chalmers University of Technology\\
        \IEEEauthorrefmark{2}Munich Research Center, Huawei Technologies Duesseldorf GmbH\\
        \IEEEauthorrefmark{3}Department of Telecommunications and Systems Engineering, Universitat Autonoma de Barcelona\\
        \IEEEauthorrefmark{4}Department of Electrical and Computer Engineering, Technische Universit\"at M\"unchen
        }

\maketitle

\begin{abstract}
The application of cooperative localization in vehicular networks is attractive to improve accuracy and coverage. Conventional distance measurements between vehicles are limited by the need for synchronization and provide no heading information of the vehicle. To address this, we present a cooperative localization algorithm using \ac{PLBP} utilizing \ac{AoA}-only measurements. Simulation results show that both directional and positional \ac{RMSE} of vehicles can be decreased significantly and converge to a low value in a few iterations. Furthermore, the influence of parameters for the vehicular network, such as vehicle density, communication radius, prior uncertainty and \ac{AoA} measurements noise, is analyzed. 
\end{abstract}

\IEEEpeerreviewmaketitle

\acresetall

\section{Introduction}

Vehicular localization with high precision is of great importance for future autonomous driving. Among different possibilities, e.g., \ac{GNSS}~\cite{gleason2009gnss}, cooperative localization~\cite{wymeersch2009cooperative} enables the possibility for \ac{MP} between vehicles, which can lead to more accurate positioning and increased positioning coverage. In cooperative localization, vehicles use on-board sensors, including 5G front-end, radar and stereo cameras~\cite{de2017survey}, to obtain measurements relative to the positions of nearby vehicles. Vehicles exchange information related to relative positions and own position estimates to obtain an approximation of their own posterior distribution. \ac{BP}~\cite{kschischang2001factor} is a well-known framework for Bayesian inference that can be applied for the cooperative localization problem~\cite{wymeersch2009cooperative}. 
Cooperative localization is particularly advantageous when vehicles have different prior localization accuracy, because vehicles with high-quality sensors can help vehicles with low quality sensors to reduce their localization errors. The last point is practical in the foreseeable future because vehicles with different levels of sensing precision are expected to coexist\cite{steinmetz2019theoretical}. 
\begin{figure}
    \centering
    \includegraphics[width = 1\linewidth]{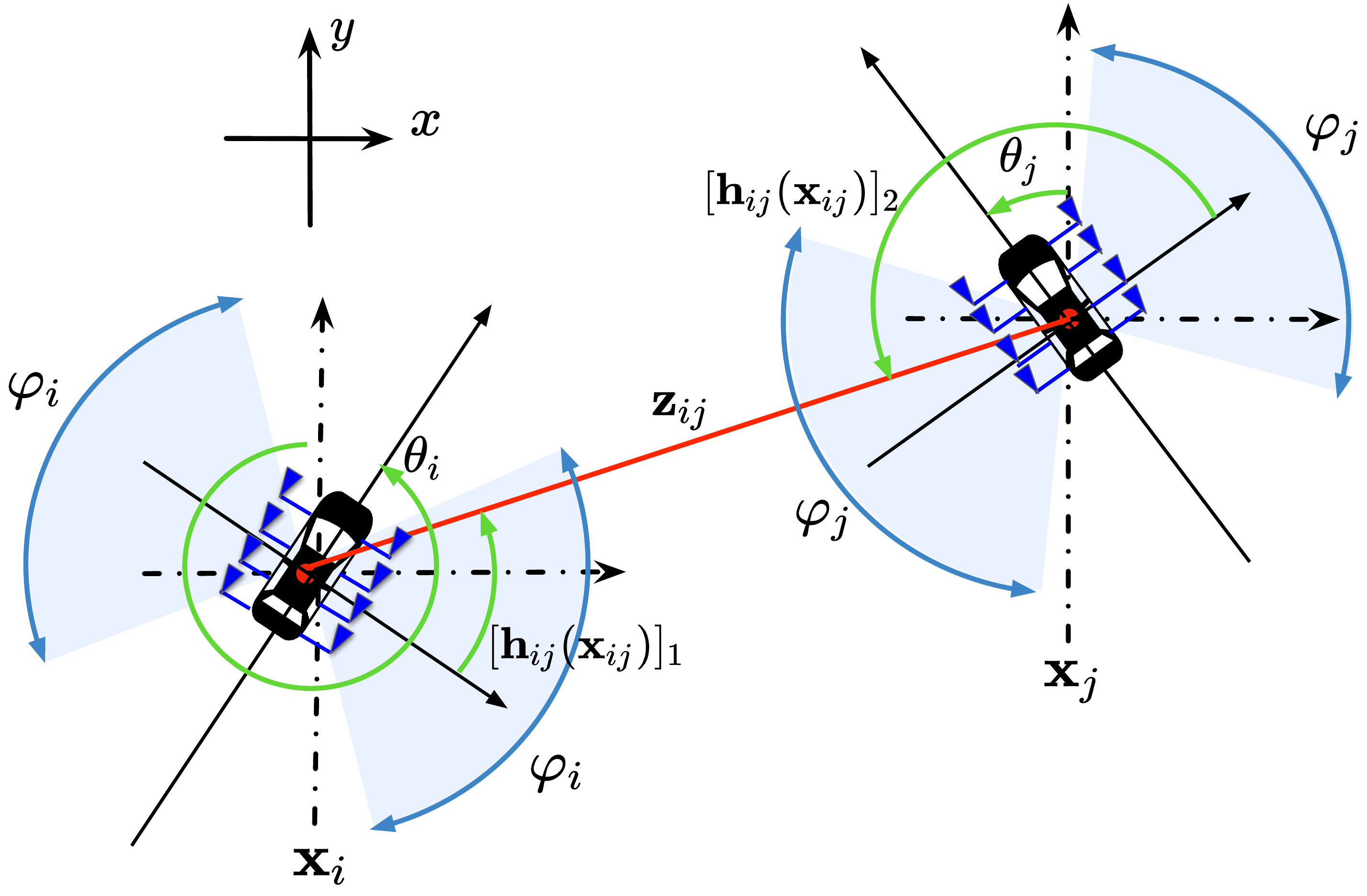}
    \caption{Geometric model of two vehicles. Vehicle $i$ measures the \ac{AoA} $[\mathbf{h}_{ij}(\mathbf{x}_{ij})]_{1}$ from vehicle $j$, and vehicle $j$ measures $[\mathbf{h}_{ij}(\mathbf{x}_{ij})]_{2}$.}
    \label{fig:data_model}
\end{figure}

The performance of any localization system is limited by the underlying measurements.  Conventional measurements include \emph{distance} and \emph{angle} between vehicles. In terms of distance measurements, radar can provide high accuracy, but does not include identity information of the target, required for \ac{MP}. Measurements based on the travel time of radio signals (\ac{TOA} and \ac{TDOA}) can provide such identity information~\cite{mohammadabadi2014cooperative, catovic2004cramer,gholami2011hybrid}.  However, \ac{TOA} and \ac{TDOA} are challenged by the synchronization requirements~\cite{catovic2004cramer}. The clocks of two vehicles need to be synchronized such that the delay can be computed. This can lead to significant localization error because of small clock error~\cite{buehrer2018collaborative}, or to use two-way \ac{TOA} with round-trip delay time instead of the one-way delay to avoid synchronization, which doubles the resource requirement. Achieving a ranging accuracy lower than 10 m by \ac{TOA}/\ac{TDOA} is very challenging  in vehicular environments~\cite{alam2013cooperative}. In contrast, \ac{AoA} is readily available when the receiver is equipped with an antenna array~\cite{sakagami1992vehicle, fascista2017angle, kakkavas2018multi, garcia2019gaussian}: \cite{kakkavas2018multi} has investigated the performance of \ac{V2V} relative positioning using \ac{AoA} measurements from multiple receiving arrays on the vehicle, and the achieved positioning accuracy met requirements of 5G \ac{NR} \ac{V2X} standardization. While \ac{AoA} measurements are attractive from a practical point of view, the integration in \ac{MP} is non-trivial. Due to the nonlinear relation between the \ac{AoA} and the vehicle state, analytical computation of the messages in \ac{BP} is not possible. Approximations include the use of particles~\cite{etzlinger2013cooperative,savic2013cooperative} or linearization of the measurement model~\cite{wan2000unscented}.  While the increasing number of particles gives better approximation performance, it also increases the computation complexity. To address this problem, \cite{garcia2019gaussian} uses a \ac{VMF} model for the measurement likelihood and performs \ac{PLBP} \cite{garcia2015posterior}, for a scenario with unknown positions but known orientation. 


In this paper, we consider a cooperative localization problem where vehicles' positions and orientations are unknown. We apply Gaussian parametric BP~\cite{yuan2016cooperative} for the MP, which reduces the communication resource overhead and computational complexity compared to a particle approach. To pass those messages through the nonlinear angle measurement model, \ac{PL}~\cite{garcia2015posterior} is applied to linearize the model using \ac{SLR} with respect to the posterior, which can be calculated by the current messages~\cite{garcia2018cooperative}. Based on the linearized model, the BP is then performed to update the new beliefs. This \ac{PLBP} procedure can be iterated so that the posterior \ac{PDF} of the vehicle position and orientation can converge. 

\section{Problem Statement}\label{section:prob_state}
We consider a network comprising a set of vehicles $\mathcal{V} = \{1,...,N\}$. A set of communication links $\mathcal{E} \subset \mathcal{V}\times \mathcal{V}$ are considered to connect each vehicle according to a communication radius $r$. The neighbor set of vehicle $i$ is denoted by $\mathcal{N}_i$. Each vehicle $i \in \mathcal{V}$ has a state $\mathbf{x}_i \in \mathbb{R}^3$, comprising the 2D position $[x_i,y_i]^{\mathsf{T}}$ and the heading $\theta_i \in (-\pi,\pi]$.  We denote the joint state of vehicles $i$ and $j$ as $\mathbf{x}_{ij}=[\mathbf{x}^{\mathsf{T}}_i \mathbf{x}^{\mathsf{T}}_j]^{\mathsf{T}}$. Each vehicle is assumed to have knowledge of its prior state by some accessible positioning techniques, e.g., GNSS, 
assumed to be a Gaussian density
\begin{align}
p_{i}(\mathbf{x}_{i}) = \mathcal{N}(\mathbf{x}_{i};\bm{\mu}_{i},\mathbf{P}_{i}),
\label{eq:data_distribution}
\end{align}
where $\mathcal{N}(\mathbf{x}_{i};\bm{\mu}_{i},\mathbf{P}_{i})$ denotes a Gaussian distribution in variable $\mathbf{x}_{i}$ with mean vector $\bm{\mu}_{i}=[\mu_{x}, \mu_{y}, \mu_{\theta}]^{\intercal}$ and  covariance matrix $\mathbf{P}_{i}$. The measurement model between two vehicles is shown in Fig.~\ref{fig:data_model}. Each vehicle $i$ is equipped with linear arrays on its two sides, each of which provides a \ac{FOV} $\varphi _{i}$ with $0 < \varphi _{i}\leq \pi$. Signals with an \ac{AoA} measurements within the \ac{FOV} of node can be measured.  The \ac{AoA} measurement vector $\mathbf{z}_{ij}$ between vehicles $i$ and $j$ is defined as a function of $\mathbf{x}_i$ and $\mathbf{x}_{j}$ with additive Gaussian noise
\begin{align}
    \mathbf{z}_{ij} = \mathbf{h}_{ij}(\mathbf{x}_{ij}) + \bm{\eta}_{ij},
    \label{eq:eq_measure_model}
\end{align}
where $\bm{\eta}_{ij}$ represents the measurement noise, modeled as $\bm{\eta}_{ij} \sim \mathcal{N}(\mathbf{0},\mathbf{R}_{ij})$ and $\mathbf{h}_{ij}(\mathbf{x}_{ij})$  is defined as\footnote{For simplicity we consider the center points of the two arrays on each vehicle to coincide. The effect of the relative position and orientation of the antenna arrays is outside the scope of this paper and related work can be found in~\cite{shen2010accuracy}. }
\begin{align}
\mathbf{h}_{ij}(\mathbf{x}_{ij}) = \left[\begin{array}{c}
\atantwo\left((y_j-y_i), (x_j-x_i)\right) - \theta_i\\
\atantwo((y_i-y_j), (x_i-x_j)) - \theta_j
\end{array}\right],
\label{eq:True_measure_model}
\end{align}
in which $\atantwo(y,x)$ calculate the four-quadrant inverse tangent of $y$ and $x$. However, the $\atantwo$ introduces problems because of its discontinuity at the negative semi-axis of $x$, i.e. $(x,0):x<0$. Instead of modeling the angular measurements by \ac{VMF} distribution, as \cite{garcia2019gaussian} has done, we adopt a simple ad-hoc correction from \cite{crouse2015cubature}, which is described in Appendix \ref{appendix: SLR}. We denote the vector of all measurements by $\mathbf{z}=[\mathbf{z}_{ij}]_{i,j\in \mathcal{N}_i}$ and the vector of all vehicles' states by $\mathbf{x}$. The goal of the network is to compute $p_i(\mathbf{x}_i |\mathbf{z})$, for each vehicle. 

\section{Belief Propagation and Posterior Linearization}
\subsection{Belief Propagation Formulation}
The standard approach to solve the localization problem is to use belief propagation. We first factorize the joint \ac{PDF} 
\begin{align}
    p(\mathbf{x} ,\mathbf{z}) & =p(\mathbf{x})p(\mathbf{z} |\mathbf{x})\\
    & = \prod_{i=1}^{N}p_i(\mathbf{x}_i) \prod_{j \in \mathcal{N}_i, j>i} p(\mathbf{z}_{ij}|\mathbf{x}_{ij}).
\end{align}
A factor graph representation of this joint \ac{PDF} in combination with loopy \ac{BP} allows the computation of approximations of the marginal posteriors $p_i(\mathbf{x}_i |\mathbf{z})$. The \ac{BP} message passing rules at iteration $k$ are as follows (assuming $j \in \mathcal{N}_i$)\cite{kschischang2001factor}
\begin{align}
b_j^{(k-1)}(\mathbf{x}_{j})&  \propto p_j(\mathbf{x}_{j}) \prod_{i \in \mathcal{N}_j}  m^{(k-1)}_{i \to j}(\mathbf{x}_j)\label{eq:MP1} \\ 
     m^{(k)}_{j \to i}(\mathbf{x}_i) & \propto  \int p(\mathbf{z}_{ij}|\mathbf{x}_{ij}) \frac{b_j^{(k-1)}(\mathbf{x}_{j})}{m^{(k-1)}_{i \to j}(\mathbf{x}_j)}\text{d}\mathbf{x}_{j}. \label{eq:MP2}
\end{align}
The approximate marginal posterior at iteration $k$ is $p_j(\mathbf{x}_j |\mathbf{z}) \approx b_j^{(k)}(\mathbf{x}_{j})$. The process is initialized at $k=0$ by $b_j^{(0)}(\mathbf{x}_{j})=p_j(\mathbf{x}_{j})$ and $m^{(0)}_{i \to j}(\mathbf{x}_j)=1$. 
The joint posterior of $\mathbf{x}_{i},\mathbf{x}_{j}$ can also  be approximated by\cite{kschischang2001factor}
\begin{align}
b^{(k)}(\mathbf{x}_{ij}) &\propto
p(\mathbf{z}_{ij}|\mathbf{x}_{ij}) \frac{b_i^{(k)}(\mathbf{x}_{i})b_j^{(k)}(\mathbf{x}_{j})}{m^{(k)}_{i \to j}(\mathbf{x}_j){m^{(k)}_{j \to i}(\mathbf{x}_i)}}. \label{eq:joint_posterior}
\end{align}
However, due to the nonlinear observation model \eqref{eq:eq_measure_model}, in general \ac{BP} cannot be executed in closed form: neither the integral \eqref{eq:MP1} nor the product \eqref{eq:MP2} can be computed exactly, except when the observation model is linear with Gaussian noise \cite{garcia2015posterior}. 
This motivates the following linearization procedure. 

\subsection{Linearization}\label{section:linearization}
\begin{figure}[ht]
     \centering
     \includegraphics[width=0.9\linewidth]{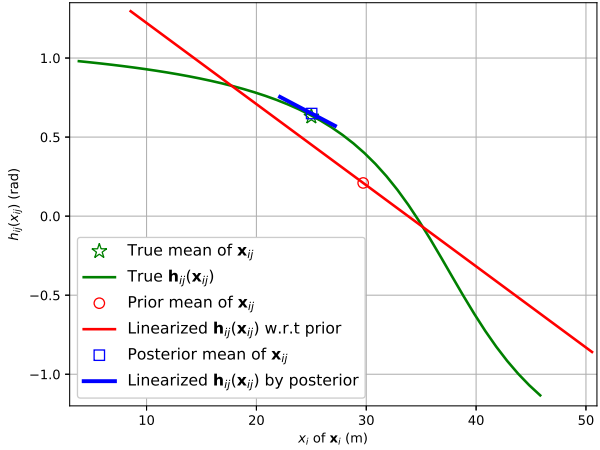}
     \caption{The true measurement model $\mathbf{h}_{ij}(\mathbf{x}_{ij})$ and its approximations by \ac{SLR} with respect to the prior and posterior, as a function of the $x$-dimension of $\mathbf{x}_i$. The length of the red and blue lines represent 2 standard deviations of the prior and posterior linearized models, respectively.
     }
     \label{fig:linearization-performance}
 \end{figure}

Given a belief $b^{(k)}(\mathbf{x}_{ij})$, we approximate the observation model by 
\begin{align}
    \mathbf{h}_{ij}(\mathbf{x}_{ij}) & \approx 
    \mathbf{C}_{ij}\tilde{\mathbf{x}}_{ij}+\mathbf{e}_{ij}, \label{eq:linearizedMeasurements}
\end{align}
where 
$\mathbf{e}_{ij}\sim \mathcal{N}(\mathbf{0},\bm{\Omega}_{i,j})$,
and $\tilde{\mathbf{x}}_{ij}=[{\mathbf{x}}^{\mathsf{T}}_{ij}~ \mathbf{1}^{\mathsf{T}}]^{\mathsf{T}}$. $\mathbf{C}_{ij}$ is selected to minimize the \ac{MSE} over the given joint belief $b^{(k)}(\mathbf{x}_{ij})$: 
\begin{align}
    \arg \underset{\mathbf{C}_{ij}}{\min} &~ \mathbb{E} \{  \Vert \mathbf{h}_{ij}(\mathbf{x}_{ij})-\mathbf{C}_{ij}\tilde{\mathbf{x}}_{ij}\Vert ^2\}. \label{eq:optimizationProblem}
\end{align}
Once $\mathbf{C}_{ij}$ is determined, we find that $\bm{\Omega}_{i,j} = \Vert \mathbf{h}_{ij}(\mathbf{x}_{ij})-\mathbf{C}_{ij}\tilde{\mathbf{x}}_{ij}\Vert^2$.
To solve this optimization problem, the \ac{SLR}~\cite{garcia2015posterior} with respect to the posterior \ac{PDF} is performed, where the details are presented in Appendix \ref{appendix: SLR}. To visualize the advantage of posterior \ac{SLR}, Fig.~\ref{fig:linearization-performance} shows the true measurement model (\ref{eq:True_measure_model}) and its approximations (\ref{eq:linearizedMeasurements}) with respect to posterior and prior. We observe that the linearized model by posterior \ac{SLR} is more accurate and has less uncertainty than the model linearized by prior \ac{SLR}.

\subsection{Belief Propagation with Linearized Measurement Models}

Once a linearization of all measurement models is obtained, \ac{BP} is performed as follows. The likelihood function is now of the form 
\begin{align}
    &  p(\mathbf{z}_{ij}|\mathbf{x}_{ij}) \propto \label{eq:linearizedlikelihood} \\
    & \exp\left(-\frac{1}{2}(\mathbf{z}_{ij}-\mathbf{C}_{ij}\tilde{\mathbf{x}}_{ij})^{\mathsf{T}} \bm{\Sigma}^{-1}_{ij} (\mathbf{z}_{ij}-\mathbf{C}_{ij}\tilde{\mathbf{x}}_{ij})\right), \nonumber 
\end{align}
where $\bm{\Sigma}_{ij}=\bm{\Omega}_{ij}+\mathbf{R}_{ij}$. 
This formulation now allows closed-form Gaussian message passing according to \eqref{eq:MP1}--\eqref{eq:MP2} and \eqref{eq:joint_posterior}. The details of the implementation are provided in the Appendix \ref{appendix: BP}.

The overall algorithm thus operates as described in Algorithm \ref{alg:algorithm 1}. The algorithm requires a selection of $K$ (the number of linearization iterations) and $M$ (the number of BP iterations per linearization step). The overall complexity per vehicle is approximately $\mathcal{O}(KM\bar{N}D^3)$, where $D$ is the state dimension and $\bar{N}$ is the average number of neighbors.
\begin{algorithm}
\caption{: Iterative Cooperative Localization}
\begin{algorithmic}
\FOR{$k = 1$ to $K$}
\STATE Given the current beliefs $b^{(k-1)}(\mathbf{x}_{ij})$, solve \eqref{eq:optimizationProblem} for each $(i,j) \in \mathcal{E}$ to obtain \eqref{eq:linearizedlikelihood}.
\STATE Run $M$ iterations of BP on the linearized model.
\STATE Compute joint beliefs $b^{(k)}(\mathbf{x}_{ij})$ at the current BP iteration. 
\ENDFOR
\STATE Return marginal beliefs.
 \end{algorithmic}
  \label{alg:algorithm 1}
 \end{algorithm}
\section{Simulation Results}
In this section we simulated a vehicular network scenario and analyzed the performance of the designed Algorithm \ref{alg:algorithm 1}. First, the localization and orientation performance of Algorithm \ref{alg:algorithm 1} in the vehicle network is evaluated by the positional and directional \ac{RMSE}. Then, based on this scenario, we analyzed the impact of different network parameters. 

\subsection{Simulation Scenario}
\begin{figure}
    \centering
    \includegraphics[width=1\linewidth]{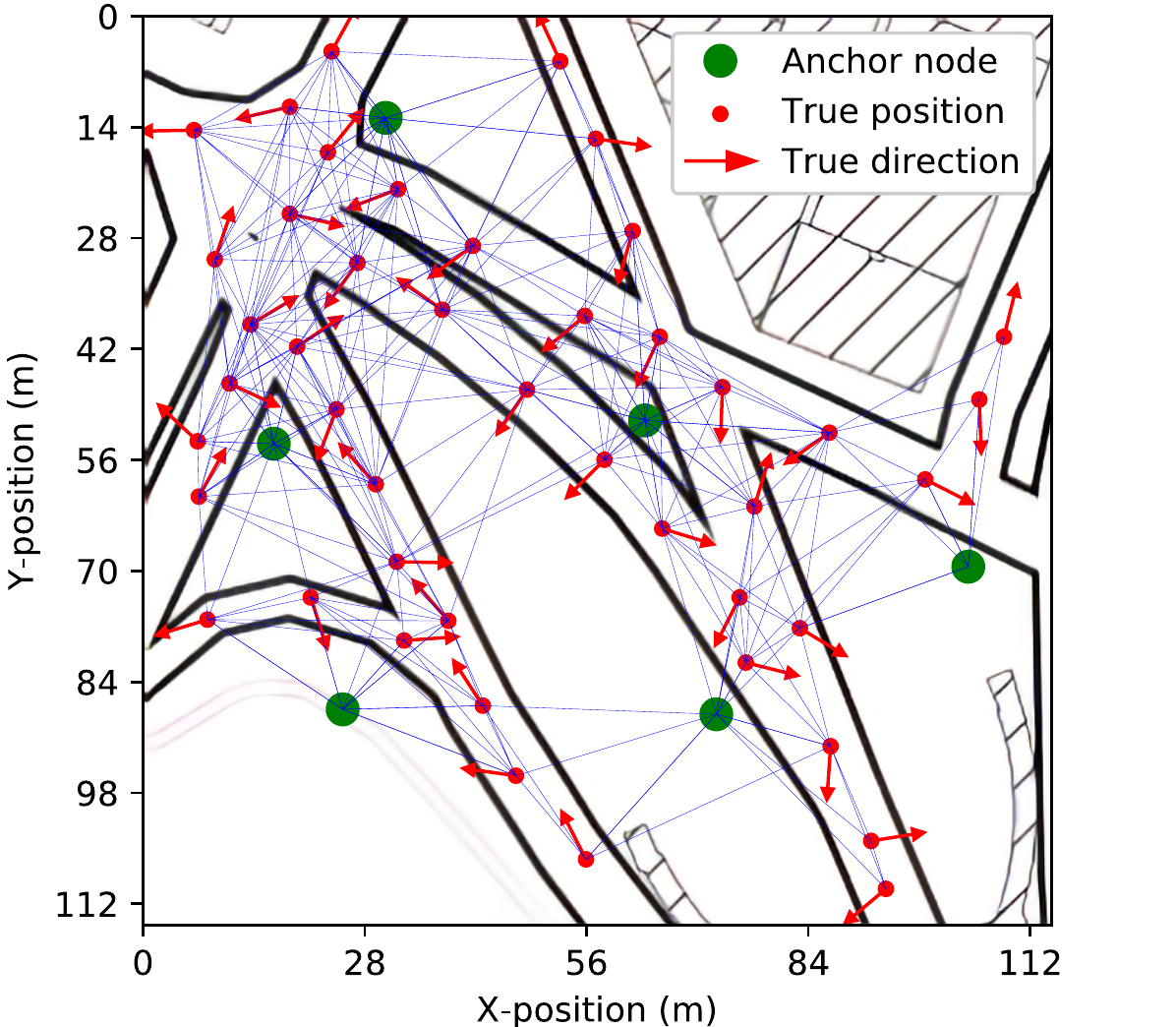}
    \caption{Scenario of the vehicular network. The interactive web map can be found in \cite{map-simple-1}.}
    \label{fig:scenario_final}
\end{figure}

The vehicular scenario is based on a road map in central New York Manhattan (latitude: $40.71590$ and longitude: $-73.99560$). The map data is generated from Stamen Map \cite{ManhattanMap} at a zoom level of 18. Within this map, the scenario is shown in Fig.~\ref{fig:scenario_final}, where 51 vehicles are possibly connected within the communication radius ($r=30~\text{m}$). The priors are set to $\mathbf{P}_{i} = \text{diag}(\sigma^{2}_{x}, \sigma^{2}_{y}, \sigma^{2}_{\theta})$. Among the vehicles, 6 are chosen as anchors (vehicles or road side units with a very concentrated prior density, set to $\text{diag}(\sigma_{x}^{2},\sigma_{y}^{2} , \sigma_{\theta}^{2})=\text{diag}(0.01, 0.01, 0.01)$). 
The interactive web map is also provided\footnote{The results of the scenario can be visualized by an interactive web map in \cite{map-simple-2}, where the red, blue, and green dots represent the true, prior and estimated positions, respectively.} \cite{map-simple-1}. The remaining parameters of this scenario are illustrated in Table \ref{tb:basic_setup}, where  $R$ denotes the constant value  of the measurement variance (approximately 18 degrees standard deviation). 
\begin{table}
\centering
\caption{Setup parameters for the vehicular scenario.}
\begin{tabular}{@{}llllll@{}}
\toprule
$r$ [m] &$\varphi$ [rad] & $\sigma_{x}$ [m] & $\sigma_{y}$ [m]& $\sigma_{\theta}$ [rad] & $R$ [rad$^{2}$] \\ \midrule
30&   $\pi$  & 5 & 5 & 0.35 & 0.10 \\ \bottomrule
\end{tabular}
\label{tb:basic_setup}
\end{table}

\subsection{Results and Discussion}

\subsubsection{Convergence Speed}
\begin{figure}
    \includegraphics[width=1\linewidth]{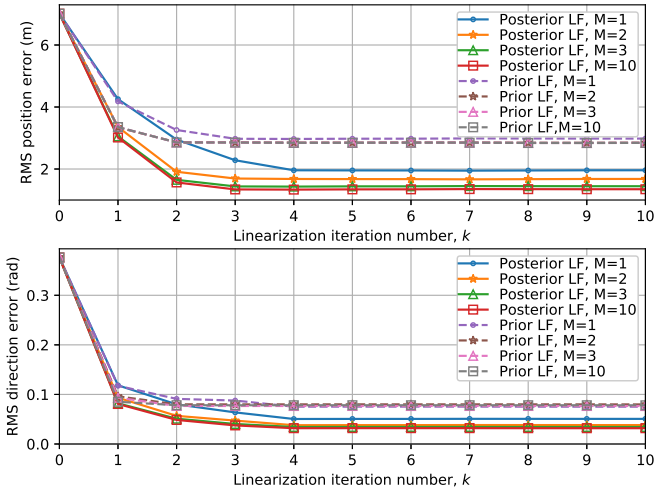}
    \caption{RMS position and direction error against the number of linearization iteration $k$. The initial position and direction RMSE of vehicles are 7.01m and 0.38 rad, respectively.}
    \label{fig:RMSE_P_D}
\end{figure}
In order to examine the performance of Algorithm \ref{alg:algorithm 1}, in Fig. \ref{fig:RMSE_P_D} we plot the RMS position and direction error against the number of linearization iteration $K$. Notice the performance gap between the prior \ac{LF} (dotted lines) and the posterior \ac{LF} (solid lines). After each belief propagation iteration, the posterior of each vehicle is closer to the true state than the prior, so the belief propagation has a better performance on the posterior linearization measurement model. Both position RMSE and direction RMSE converged for linearization iteration number larger than 4. Meanwhile, increasing $M$ from 1 to 3 provides significant improvements for both position and orientation estimation accuracy as the beliefs are more accurate. The improvement becomes very small for $M$ greater than 3. 

\subsubsection{Localization Performance}
\begin{figure}
    \centering
    \includegraphics[width=1\linewidth]{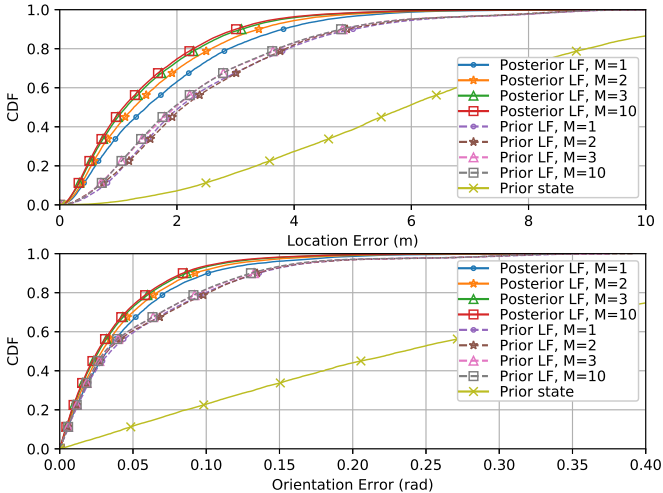}
    \caption{CDF of localization and orientation error, $K=10$.}
    \label{fig:CDF_P_D}
\end{figure}

 While the above results show the average RMSE of the position and direction, Fig.~\ref{fig:CDF_P_D} shows the cumulative distribution functions (CDFs) of the position and direction errors for $K=10$ for different values of $M$.  We observe that for $M=3$ the performance is similar to $M=10$ and that nearly all vehicles can be localized with a position error less then 4 meters and an orientation error less than 0.15 radians (8 degrees). The importance of posterior linearization over prior linearization is again clear.  
\subsubsection{Impact of Network parameters}
\begin{figure}
    \centering
    \includegraphics[width=1\linewidth]{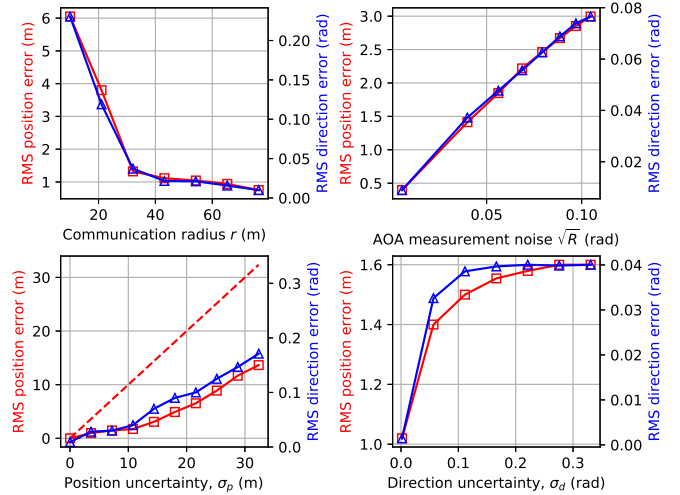}
    \caption{The impact of 4 vehicle network parameters on localization and orientation performance. Lines with square and triangle markers represent the position and orientation RMSE, respectively. $K=10$, $M=10$ and posterior LF are applied.}
    \label{fig:impact-parameters}
\end{figure}
Here, we analyze the impact of modifying the scenario parameters in Table \ref{tb:basic_setup} on localization and orientation estimation performance. In Fig.~\ref{fig:impact-parameters}, we evaluate 4 parameters separately, namely communication radius ($r$), measurement noise variance ($R$), prior uncertainty in position ($\sigma_{p} = (\sigma_{x}^{2}+ \sigma_{y}^{2})^{1/2}$) and prior uncertainty in orientation ($\sigma_{\theta}$) in 4 sub-figures, by plotting the position and direction RMSE as functions of one of them, while keeping the rest fixed to the values of Table \ref{tb:basic_setup}.
\begin{itemize}[noitemsep,topsep=-10pt]
    \item The top left sub-figure shows the impact of the communication radius $r$. Both RMSEs are reduced rapidly by increasing $r$ from 10 m to 30 m since each vehicle has more neighbors and the network connectivity increases quickly, up to the point where all vehicles are in each others' communication range. We note that with increased connectivity comes increased computational complexity. 
    \item In the top right sub-figure, we vary the \ac{AoA} measurements noise variance $R$. We note that both direction and position RMSE increase approximately linearly in $\sqrt{R}$. This emphasizes the need for good measurements. 
\item The influence of the prior position uncertainty ($\sqrt{\sigma_{x}, \sigma_{y}}$)  is shown in the bottom left sub-figure. The red dashed line describes the prior position RMSE. We notice the increase of $\sigma_p$ from 0 m to 10 m  has small effect on both position and direction performance (less than 2 m/0.05 rad), showing the good performance of the proposed method. For position uncertainty over 10 m, Algorithm \ref{alg:algorithm 1} is still able to improve performance over the prior RMSE, but leads to progressively larger errors. This is in contrast to range-based cooperative localization \cite{wymeersch2009cooperative}, where no prior information was needed. 
\item The influence of the direction uncertainty ($\sigma_{\theta}$) is shown in the bottom right sub-figure, where we observe a rapid increase in RMSE. This is because the \ac{AoA} measurements depend on the orientation of the receiving vehicles. For larger prior orientation uncertainty, Algorithm \ref{alg:algorithm 1} is less affected. 
\end{itemize}

\section{Conclusion}
We have applied PLBP to cooperative localization (position and orientation estimation) of vehicles with \ac{AoA}-only measurements. 
Multiple conditions of the vehicular network, including the vehicle density, communication radius, prior uncertainty and measurement noise variance have been discussed. Numerical results show that the proposed algorithm has good performance in terms of both position and orientation estimation, and only a few iterations are required for convergence. This makes the algorithm attractive for real-time processing. 
 \section*{Acknowledgment}

This research was supported, in part, by the EU Horizon 2020 project 5GCAR (Fifth Generation Communication Automotive Research and innovation) and the Spanish Ministry of Science, Innovation and Universities under Grant TEC2017-89925-R.

\appendices 
\section{Steps of the posterior linearization}
\label{appendix: SLR}
This section illustrates the procedures of SLR on the measurement model and the approximation of the parameters ($\mathbf{C}_{ij},\mathbf{\Omega}_{i,j}$) with respect to the joint posterior \ac{PDF} $p(\mathbf{x}_{ij}|\mathbf{z}_{ij})$ = $\mathcal{N}(\mathbf{x}_{ij};\bm{\mu}_{ij};\mathbf{P}_{ij}) $.
 First, according to the joint posterior of $\mathbf{x}_{i},\mathbf{x}_{j}$, we select $L$ sigma-points $\mathcal{X}_{1},...,\mathcal{X}_{L}$ and weights $\omega_{1},...,\omega_{L}$ using a sigma-point method such as the unscented transform~\cite{julier2004unscented}.
 Then we calculate the transformed sigma points by
 \begin{align}\label{eq:sigma-points}
     \mathcal{Z}_{l} = \mathbf{h}_{ij}(\mathcal{X}_{l}) \hspace{10pt} l = 1,...,L
 \end{align}
However, as mentioned in Section \ref{section:prob_state}, the function arctan has discontinuity problem at the negative $x$ semi-axis. The sigma points transformation needs an ad-hoc modification so that the difference between angles $\mathcal{Z}_l - \mathbf{z}_{ij}$ must be bounded in $\pm \pi$. $\mathcal{Z}_{l}$ can be corrected to $\hat{\mathcal{Z}_{l}}$ by the following transformation:
\begin{align}\label{eq:ad-hoc}
    \hat{\mathcal{Z}_{l}} = \mathbf{z}_{ij} + \pi - \text{modulo}((\mathbf{z}_{ij} -\mathcal{Z}_{l})+\pi)_{2\pi}
\end{align}
where $\hat{\mathcal{Z}_{l}}$ denotes the corrected sigma point, $z_{ij}$ is the \ac{AoA} measurements and $\text{modulo}(\cdot)_{2 \pi}$ represents the modulo operation.

Introducing $\mathbf{C}_{ij}=[\mathbf{A}_{ij}~ \mathbf{b}_{ij}]$, so that 
\begin{align}
    \mathbf{h}_{ij}(\mathbf{x}_{ij}) & \approx \mathbf{A}_{ij} \mathbf{x}_{ij} + \mathbf{b}_{ij}+\mathbf{e}_{ij},
\end{align}
the solution of the approximation of $\mathbf{A}_{ij},\mathbf{b}_{ij},\Omega_{i,j}$  is
\begin{align}
 \mathbf{A}_{ij} &=  \mathbf{C}_{xz}^{\mathsf{T}}\mathbf{P}_{ij}^{-1}\label{eq:sigma-A}\\
 \mathbf{b}_{ij} &= \bar{z} -  \mathbf{A}_{ij}\bm{\mu}_{ij}\label{eq:sigma-b}\\
 \mathbf{\Omega}_{i,j}& = \mathbf{C}_{zz} - \mathbf{A}_{ij}\mathbf{P}_{ij}\mathbf{A}_{ij}^{\mathsf{T}}\label{eq:sigma-Omega}
\end{align}
where $\bar{z}$, $\mathbf{C}_{xz}$ and $\mathbf{C}_{zz}$ are approximated using the sigma-points \eqref{eq:ad-hoc} and weights by 
\begin{align}
     \bar{z} \approx& \sum_{j=1}^{L}\omega_{j}\hat{\mathcal{Z}_{l}} \label{eq:sigma-z} \\
     \mathbf{C}_{xz} \approx& \sum_{j=1}^{L}\omega_{j}(\mathcal{X}_{j} - \bm{\mu}_{ij})(\hat{\mathcal{Z}_{l}} - \bar{z})^{\mathsf{T}}\label{eq:sigma-Psi}\\
     \mathbf{C}_{zz} \approx& \sum_{j=1}^{L}\omega_{j}(\hat{\mathcal{Z}_{l}} - \bar{z})(\hat{\mathcal{Z}_{l}} - \bar{z})^{\mathsf{T}}.\label{eq:sigma-Phi}
\end{align}

\section{Implementation of BP in the linearized model}\label{appendix: BP}
This section illustrates the derivation of equation \eqref{eq:MP1}--\eqref{eq:MP2}
 and \eqref{eq:joint_posterior}.
Once we have the approximated linearization model \ref{eq:linearizedMeasurements}, we can represent the BP message $m^{(k)}_{i\rightarrow j}$ by the Gaussian format~\cite{garcia2018cooperative}
\begin{equation}\label{eq:msg_a}
    m^{(k)}_{i\to j}(\mathbf{x}_j) \propto \mathcal{N}(\bm{\alpha}^{(k)}_{ij}; \mathbf{H}^{(k)}_{ij}\mathbf{x}_{j},\bm{\Gamma}^{(k)}_{ij})
\end{equation}
where $\bm{\alpha}^{(k)}_{ij}$, $\mathbf{H}^{(k)}_{ij}$ and $\bm{\Gamma}^{(k)}_{ij}$ are
\begin{subequations}\label{eq:msg_abc}
\begin{equation}\label{eq:msg_b}
    \bm{\alpha}^{(k)}_{ij} = [\mathbf{z}_{ij}]_1 - \mathbf{A}_{i}\bm{\mu}^{(k-1)}_{ij} - b_{ij}
\end{equation}
\begin{equation}\label{eq:msg_c}
    \mathbf{H}^{(k)}_{ij} = \mathbf{A}_{j}
\end{equation}
\begin{equation}\label{eq:msg_d}
\bm{\Gamma}^{(k)}_{ij} = \mathbf{R}_{ij} + \bm{\Omega}_{ij} + \mathbf{A}_{i}\mathbf{P}^{(k-1)}_{ij}\mathbf{A}_{i}^\intercal
\end{equation}
\end{subequations}
where $[\mathbf{z}_{ij}]_1$ is the \ac{AoA} measurement received by vehicle $i$, $\mathbf{A}_{i},\mathbf{A}_{j}$ are defined at Section \ref{section:linearization} and $\bm{\mu}^{(k-1)}_{ij}$ and $\mathbf{P}^{(k-1)}_{ij}$ are found from the relation
\begin{equation}
    \begin{split}
    \mathcal{N}(\bm{\mu}^{(k-1)}_{ij},\mathbf{P}^{(k-1)}_{ij}) \propto\mathcal{N}(\mathbf{x}_i;\bm{\mu}_i,\mathbf{P}_i) \prod_{j'\in \mathcal{N}_i\backslash j} m^{(k-1)}_{j'\rightarrow i}(\mathbf{x}_{i})
    \end{split}
    \label{eq:msg_x_ij}
\end{equation}
where the Kalman update step~\cite[Algorithm 1]{garcia2018cooperative} is performed to update each message $m^{(k-1)}_{j'\to i}(\mathbf{x}_{i})$ on the prior state $\mathcal{N}(\mathbf{x}_i;\bm{\mu}_i,\mathbf{P}_i)$.
 
To get the local belief (\ref{eq:MP1}) at the $k$-th iteration, we can also use Kalman filter update step to update the vehicle prior with all its incoming messages. 
\begin{equation}
\label{eq:marginal_distribution}
    b_j^{(k)}(\mathbf{x}_{j}) =  \mathcal{N}(\mathbf{x}_{j};\bm{\mu}_{j},\mathbf{P}_{j}) \times \prod_{i \in \mathcal{N}_j}  m^{(k)}_{i \to j}(\mathbf{x}_j)
\end{equation}
The $k$-th iteration joint posterior (\ref{eq:joint_posterior}) is expressed as~\cite{garcia2018cooperative}
\begin{align}
& b^{(k)}(\mathbf{x}_{ij})=\mathcal{N}(\mathbf{x}_{i};\bm{\mu}_{i},\mathbf{P}_{i}) \prod_{j'\in \mathcal{N}_i\backslash j} m^{(k)}_{j'\to i}(\mathbf{x}_{i})
\\ & \times \mathcal{N}(\mathbf{x}_{j},\bm{\mu}_{j},\mathbf{P}_{j})\times \prod_{i'\in \mathcal{N}_j\backslash i} m^{(k)}_{i'\rightarrow j}(\mathbf{x}_{j})  p(\mathbf{z}_{ij}|\mathbf{x}_{i},\mathbf{x}_{j}) \nonumber 
\end{align}
where we can also apply Kalman filter update~\cite[Algorithm 1]{garcia2018cooperative} as in (\ref{eq:msg_x_ij}).

\ifCLASSOPTIONcaptionsoff
  \newpage
\fi



%


\bibliographystyle{IEEEtran}
\bibliography{reference} \label{reference}

%

\end{document}

%% file: acronyms.tex
\acrodef{3-G}{3-Generation}
\acrodef{3GPP}{3rd Generation Partnership Project}

\acrodef{AoA}{angle-of-arrival}

\acrodef{AWGN}{Additive White Gaussian Noise} 

\acrodef{BP}{Belief propagation}

\acrodef{CDF}{cumulative distribution function}

\acrodef{DoA}{Direction-of-Arrival}

\acrodef{FIM}{Fisher information matrix}
\acrodef{FOV}{field of view}

\acrodef{GNSS}{global navigation satellite system}

\acrodef{GPS}{Global Positioning System}

\acrodef{LS}{Least Squares}

\acrodef{LF}{linearization filter}

\acrodef{MAP}{Maximum A Posteriori}
\acrodef{MMSE}{Minimum Mean Squared Error}

\acrodef{MSE}{mean square error}
\acrodef{ML}{Maximum Likelihood}
\acrodef{MP}{message passing}
\acrodef{NR}{New Radio}

\acrodef{PSD}{power spectral density}
\acrodef{PDF}{probability density function}

\acrodef{RMSE}{root mean squared error}
\acrodef{SLR}{statistical linear regression}
\acrodef{TDOA}{time-difference-of-arrival}
\acrodef{TOA}{time-of-arrival}
\acrodef{UWB}{Ultra-WideBand}
\acrodef{UHF}{Ultra High Frequency}
\acrodef{VMF}{von Mises-Fisher}
\acrodef{V2V}{vehicle-to-vehicle}
\acrodef{V2X}{vehicle-to-everything}

\acrodef{PLBP}{posterior linearization belief propagation}

\acrodef{PL}{posterior linearization}
\acrodef{SLR}{statistical linear regression}